\documentclass[twoside,11pt,final]{entics}
\usepackage{enticsmacro}
\usepackage{graphicx}
\usepackage[all]{xy}
\usepackage[utf8]{inputenc}
\usepackage{microtype}
\usepackage{enumitem}
\usepackage{mathtools}
\usepackage{wrapfig}
\usepackage{mathrsfs}
\usepackage{amsmath}
\usepackage{amssymb}
\usepackage{stmaryrd}

\usepackage{xparse}

\NewDocumentCommand\mprod{e_}{
  \IfNoValueTF{#1}
    {\prod}
    {
      \prod
    }
  }

\NewDocumentCommand\msum{e_}{
  \IfNoValueTF{#1}
    {\sum}
    {
      \sum
    }
  }
  
\usepackage{hyperref}
\hypersetup{
    colorlinks=true,
    linkcolor=blue,
    filecolor=blue,
    urlcolor=blue,
    citecolor=blue
}

\usepackage{tikz}
\usetikzlibrary{fit}
\usetikzlibrary{backgrounds, fit, positioning}
\usetikzlibrary{decorations.pathmorphing,shapes}

\usepackage{mathpartir}

\newcommand\pholder{(\text{-})}
\newcommand\spholder{\text{-}}
\newcommand\op{\mathsf{op}}
\newcommand\catvar[1]{\mathscr{#1}}
\newcommand\catname[1]{\mathbf{#1}}
\renewcommand\AA{\catvar{A}}

\newcommand\CC{\catvar{M}}
\newcommand\DD{\catvar{D}}
\newcommand\Set{\catname{Set}}
\newcommand\Mon{\catname{Mon}}
\newcommand\DDX{X}
\newcommand\DDY{Y}
\newcommand\AAA{A}
\newcommand\AAB{B}
\newcommand\AAC{C}
\newcommand{\id}{\mathsf{id}}
\newcommand{\inl}{\mathsf{inl}}
\newcommand{\inr}{\mathsf{inr}}

\newcommand{\Id}{\mathsf{Id}}

\newcommand\Alg{\text{-}\mathrm{Alg}}
\newcommand\contname[1]{\mathfrak C^{#1}}

\newcommand\bdin{\mathbf{SDin}}

\newcommand\repname{\mathbb R}
\newcommand\rep[2]{{\repname #1}}
\newcommand\repxx{\rep{X}{X}}
\newcommand\urepname{R}
\newcommand\urep[2]{{\urepname #1 #2}}
\newcommand\urepxx{\urep{X}{X}}
\newcommand\urepyy{\urep{Y}{Y}}
\newcommand{\appm}{\mathsf{run}}

\newcommand{\ufreem}[1]{\widehat{#1}}

\newcommand{\tocay}[1]{\lfloor #1 \rfloor}
\newcommand{\tobdn}[1]{\lceil #1 \rceil}

\newcommand{\run}{\mathsf{run}}

\newcommand{\CR}{\contname{R}}

\newcommand{\power}{\pitchfork}
\newcommand{\Ubar}{\overline{U}}

\newcommand{\flift}[1]{\widetilde{#1}}
\newcommand{\cod}[1]{\mathfrak{T}^{#1}}
\newcommand{\MM}{T}
\newcommand{\MMA}{TA}
\newcommand{\MMB}{TB}

\def\conf{MFPS 2025} 	
\volume{5}			
\def\volu{5}			
\def\papno{17}			
\def\mydoi{16689}
\def\runtitle{Strong dinatural transformations\ldots}
\def\copynames{M. Piróg, F. Sieczkowski}
\def\lastname{Piróg, Sieczkowski}
\begin{document}
\begin{frontmatter}
  \title{Strong Dinatural Transformations and\\ Generalised Codensity Monads}
  \author{Maciej Piróg\thanksref{a}}	
  \address{Institute of Computer Science\\ University of Wrocław\\			
    Wrocław, Poland}  							
   \thanks[a]{Email: \href{mailto:mpirog@cs.uni.wroc.pl} {\texttt{\normalshape
        mpirog@cs.uni.wroc.pl}}} 
  \author{Filip Sieczkowski\thanksref{b}}
  \address{School of Mathematical \& Computer Sciences\\Heriot-Watt University\\
    Edinburgh, Scotland} 
  \thanks[b]{Email:  \href{mailto:f.sieczkowski@hw.ac.uk} {\texttt{\normalshape
        f.sieczkowski@hw.ac.uk}}}
\begin{abstract}
  We introduce \emph{dicodensity monads}: a generalisation of pointwise
  codensity monads generated by functors to monads generated by mixed-variant
  bifunctors. Our construction is based on the notion of strong dinaturality
  (also known as Barr dinaturality), and is inspired by denotational models of
  certain types in polymorphic lambda calculi -- in particular, a form of
  continuation monads with universally quantified variables, such as the Church
  encoding of the list monad in System~F. Extending some previous results on
  Cayley-style representations, we provide a set of sufficient conditions to
  establish an isomorphism between a monad and the dicodensity monad for a given
  bifunctor. Then, we focus on the class of monads obtained by instantiating our
  construction with hom-functors and, more generally, bifunctors given by
  objects of homomorphisms (that is, internalised hom-sets between
  Eilenberg--Moore algebras). This gives us, for example, novel presentations of
  monads generated by different kinds of semirings and other theories used to
  model ordered nondeterministic computations.
\end{abstract}
\end{frontmatter}

\section{Introduction}

Consider the continuation monad for some fixed answer $O$, given at $A$ as:
\begin{equation}\label{eq:dummyLabel}
(A \to O) \to O
\end{equation}
This can be read as a type in a type theory (in which case both $A$ and $O$ are types) or categorically, as an object in a category $\AA$, where it denotes the $\AA[A,O]$-fold product of~$O$ (a power). Continuation monads occupy a special place in the space of all monads, which can be seen through their possible generalisations in different settings.

One such generalisation, on the categorical side, is the codensity monad~\cite{kockCodensity} on a category~$\AA$ induced by a functor $G : \DD \to \AA$, which we denote $\cod{G}$. We discuss this monad in general in Section~\ref{s:cod}, but to see its connection with the continuation monad, it suffices to say that if $\AA= \Set$, then $\cod{G}A$ is given by the set of transformations natural in~$X$ of the following type:
\begin{equation}\label{eq:anotherDummyLabel}
{(GX)}^{A} \to GX
\end{equation}
Broadly speaking, the codensity monad $\cod{G}$ approximates the monad $GF$ generated by a pair of adjoint functors, even if $G$'s left adjoint~$F$ does not exist (see~\cite{ADAMEK2021107486,avery,leinster,Sipos} for examples). Moreover, if~$G$ does have a left adjoint $F$, the monad $\cod{G}$ is isomorphic to~$GF$. Thus, codensity monads give alternative presentations of monads, which is used, for example, in functional programming to boost performance of monadic computations~\cite{DBLP:conf/mpc/Hinze12,DBLP:conf/mpc/Voigtlander08,DBLP:conf/mpc/WuS15}.

On the side of type theory, say in System~F, one can always give a monadic structure
(with laws satisfied up to $\beta\eta$-equivalence) to the following family of types varying in~$A$, where~$T$ is any type that can contain the variable~$X$:
\begin{equation}\label{eq:cayleyM}
\forall X. (A \to T) \to T
\end{equation}
In particular, $X$ can have both positive and negative occurrences in~$T$, which makes~$T$ -- at least when only the basic type formers are involved -- a mixed-variant bifunctor. One example of such a type is the Church encoding of the list monad, given by the type $\forall X. (A \to X \to X) \to X \to X$.

This motivates the notion of \emph{dicodensity monad} introduced in~Section~\ref{s:cont}: a monad~$\CR$ that generalises codensity monad to a monad generated by a mixed-variant bifunctor $R : \DD^\op \times \DD \to \AA$. The definition of~$\CR$ uses the concept of \emph{strong dinaturality}~\cite{grayIntegration} (also known as \emph{Barr dinaturality}, detailed in Section~\ref{s:pan}), which is one of a few possible generalisations of the usual notion of natural transformation between functors to the case of mixed-variant bifunctors. On $\Set$, given a bifunctor $R : \DD^\op \times \DD \to \Set$, we define $\CR A$ to be the set of transformations strongly dinatural in~$X$ of the following type:
\begin{equation}\label{eq:crSet}
(RXX)^A \to RXX
\end{equation}
On an arbitrary category, $\CR$ is given by particular \emph{objects} of strong dinatural transformations defined via a universal property.

Our construction generalises the codensity monad of a functor $G : \DD \to \AA$ in two aspects. Not only $\CR$ instantiates to $\cod{G}$ with $RXY = GY$ (Section~\ref{s:cod}), but it also enojys a similar construction as a limit over an appropriately generalised diagram (Section~\ref{sec:genCod}).
This construction also generalises our previous results~\cite{DBLP:conf/fossacs/PirogPS19}, where we studied computational effects captured by $\Set$-monads given by the formula~\eqref{eq:crSet} for bifunctors $R : \Set^\op \times \Set \to \Set$ of the shape $RXY = PX \to Y$ for a polynomial functor $P$.

For a particular bifunctor $R : \DD^\op \times \DD \to \AA$, two questions are of interest. First, we consider whether~$\CR$ exists, which may not be obvious --- for instance, for $\AA = \Set$, the collection of transformations~\eqref{eq:crSet} needs to be small enough to form a set for every $A$. Second, we may ask if $\CR$ has a more direct characterisation, specifically whether it is isomorphic to a given particular monad. Just as in the case of codensity monads, answers to these questions are not always trivial.

In Section~\ref{s:cayley}, we give a set of sufficient conditions for $\CR$ to exist and be isomorphic to the monad $UF$ given by a pair of adjoint functors. Roughly speaking, these conditions state that the bifunctor~$R$ carries a structure that is a form of representation of objects from the codomain of~$F$ in the base catgory $\AA$. For example, the monad $\mathfrak{C}^{\Set[{-}, {=}]}$ generated on $\Set$ by the hom-functor $\Set[{-}, {=}]$ is isomorphic to the list monad (the free monad of the theory of monoids), while $\Set[A,A]$ (the set of all functions $A \to A$) together with composition and identity is the usual Cayley representation of monoids with carrier $A$. The sufficient conditions additionally require the abstraction and representation morphisms to be strongly dinatural, and to appropriately commute with strong dinaturals.

Finally, in Section~\ref{s:homo}, we consider a particular class of bifunctors that encompasses hom-functors (as in the example above), for which we simplify the construction of the previous section. This is achieved by generalising hom-functors to bifunctors given by objects of homomorphisms~\cite{Kock71closedcategories}, which are internalised hom-sets of Eilenberg--Moore categories. All these bifunctors are examples of ``arrow'' types on different levels of generality, for which the sufficient conditions given in Section~\ref{s:cayley} are a form of generalised Cayley representation of monoids.

\subsection{Structure and contributions}

In Section~\ref{s:pan}, we give the necessary background on strong dinatural transformations. Then, we present our main results:

\begin{itemize}
\item In Section~\ref{s:cont}, we define the monad $\CR$ induced by a mixed-variant bifunctor $R$. We show that if $R$ is constant, $\CR$ is isomorphic to the usual continuation monad,
\item In Section~\ref{s:cod}, we show that when $R$ is dummy in its contravariant variable (that is, $RXY = GY$ for a functor $G$), the monad $\CR$ is isomorphic to the pointwise codensity monad of $G$,
\item In Section~\ref{sec:genCod}, we show a construction of $\CR$ as a limit, generalising a similar construction for the codensity monad,
\item In Section~\ref{s:cayley}, we give a set of sufficient conditions for $\CR$ to be isomorphic to a given monad,
\item In Section~\ref{s:homo}, we work out a special case of the theorem from Section~\ref{s:cayley}, in which the functor $R$ is the hom-functor of a category, or, more generally, a functor given by objects of homomorphisms~\cite{Kock71closedcategories} between algebras of a commutative monad~$\MM$. In such a case, the monad $\CR$ is isomorphic to the monad that comes about from the canonical distributive law of lists over $\MM$~\cite{article}.
\end{itemize}

\subsection{Notation}

When the types are known, we often omit subscripts in natural transformations. The letter $R$ stands for a bifunctor, and we omit parentheses, i.e., we write $RXY$ for what we would usually write as $R(X,Y)$, and $R(FX)(GY)$ for $R(FX,GY)$. We use $\AA[\AAA,\AAB]$ for the hom-functor of a category $\AA$, while $\theta$ always stands for a strong dinatural transformation. We use $x \mapsto e$ to define functions. We write $A \Rightarrow B$ for exponential objects and internal homs, but $B^A$ is sometimes used to stress that the exponential lives in $\Set$. Similarly, $\cdot$ is used for composition of morphisms, but we sometimes use $\circ$ to stress that we are composing functions.

\section{Dinaturality and Strong Dinaturality}\label{s:pan}

In this section, we give the necessary background on (strong) dinatural transformations between mixed variant bifunctors, which appear in models of (restricted) parametric polymorphism. The classic generalisation of naturality to mixed-variant bifunctors is \emph{dinaturality}, introduced by Dubuc and Street~\cite{10.1007/BFb0060443}. The general definition is as follows:

\begin{definition}
  Given two bifunctors $G,H : \DD^\op\times \DD \to \AA$, a \emph{dinatural transformation} $\xi : G \to H$ is a collection of $\AA$-morphisms $\xi_X$ indexed by objects of $\DD$, such that for all morphisms $f : X \to Y$, the following diagram commutes:
  \begin{equation*}
    \begin{tikzpicture}[-latex, scale=1]
    \node (a) at (1,0) {$GYX$};
    \node (b) at (2, 1) {$GXX$};
    \node (c) at (6, 1) {$HXX$};
    \node (d) at (7, 0) {$HXY$};
    \node (bb) at (2, -1) {$GYY$};
    \node (cc) at (6, -1) {$HYY$};
    \path[arrows={-latex}, font=\scriptsize]
    (a) edge node [left,yshift=0.1em] {$GfX$} (b)
    (a) edge node [left,yshift=-0.1em] {$GYf$} (bb)
    (b) edge node [above] {$\xi_X$} (c)
    (c) edge node [right,yshift=0.1em] {$HXf$} (d)
    (bb) edge node [below] {$\xi_Y$} (cc)
    (cc) edge node [right,yshift=-0.1em] {$HfY$} (d)
 ;
    \end{tikzpicture}
  \end{equation*}
\end{definition}

Note that in the categorical definition the domain of bifunctors $G$ and $H$ can be different than their range --- this is more general than the case in System F, where we only have a single universe of types.

While this definition is often useful, it is also rather weak. Dinatural transformations do not compose in the general case, which is a property necessary for our constructions. Moreover the dinaturality and parametricity do not necessarily correspond directly. For instance, the type of Church numerals in System~F, $\forall X. (X \to X) \to (X \to X)$, is inhabited exactly by natural numbers, while the collection of dinatural transformations $\xi_X : X^X \to X^X$ in $\Set$ is a proper class (see Paré and Román~\cite{PARE199833} for a thorough discussion). One way to strengthen dinaturality is the following definition:

\begin{definition}
  Given two bifunctors $G,H : \DD^\op\times \DD \to \AA$, a \emph{strong dinatural transformation} (also called \emph{Barr}-dinatural transformation~\cite{PARE199833}) $\theta : G \to H$ is a collection of $\AA$-morphisms~$\theta_X$ indexed by objects of $\DD$, such that for all morphisms $g_1 : Z \to GXX$, $g_2 : Z \to GYY$, and $f : X \to Y$, if the inner square in the following diagram commutes, the entire diagram commutes:
  \begin{equation*}
    \begin{tikzpicture}[-latex, scale=1]
    \node (a) at (1,0) {$Z$};
    \node (b) at (2, 1) {$GXX$};
    \node (c) at (6, 1) {$HXX$};
    \node (d) at (7, 0) {$HXY$};
    \node (bb) at (2, -1) {$GYY$};
    \node (cc) at (6, -1) {$HYY$};
\node (e) at (3, 0) {$GXY$};
    \path[arrows={-latex}, font=\scriptsize]
    (a) edge node [left,yshift=0.1em] {$g_1$} (b)
    (a) edge node [left,yshift=-0.1em] {$g_2$} (bb)
    (b) edge node [above] {$\theta_X$} (c)
    (c) edge node [right,yshift=0.1em] {$HXf$} (d)
    (bb) edge node [below] {$\theta_Y$} (cc)
    (cc) edge node [right,yshift=-0.1em] {$HfY$} (d)
 (b) edge node [right,yshift=0.1em] {$GXf$} (e)
 (bb) edge node [right,yshift=-0.1em] {$GfY$} (e)
 ;
    \end{tikzpicture}
  \end{equation*}
We denote the collection of strong dinatural transformations from $G$ to $H$ as $\bdin[G,H]$.
\end{definition}

Not all System F types of the shape $\forall X. F \to G$ have this property -- Vene~\textit{et al.}~\cite{vene2006} give a set of syntactic sufficient conditions for this to be the case (broadly speaking: when the functors have no variables in double-negative positions). However, strongly dinatural transformations are well-behaved: they compose, and the collection of strongly dinatural transformations $\xi_X : X^X \to X^X$ in $\Set$ is isomorphic to the set of natural numbers (a direct proof was given by Paré and Román~\cite{PARE199833}, but it is also a simple corollary of our Example~\ref{ex:listOnSet}). So, while famously there are no set-theoretic models of System F~\cite{DBLP:conf/sdt/Reynolds84}, we can interpret a class of sufficiently simple types even in $\Set$, and, of course, generalise this to other categories.

\section{Dicodensity Monads}\label{s:cont}

Assume locally small categories $\AA$ (`ambient') and $\DD$ (`domain'). In this section, we construct the dicodensity monad~$\CR$ on~$\AA$ induced by a functor $R : \DD^\op \times \DD \to \AA$. The loose intuition is that it is the monad given type-theoretically by $\CR \AAA = \forall \DDX \in \DD. (\AAA \to R \DDX \DDX) \to R \DDX \DDX$. In $\Set$, it is given by the set of transformations $(R \DDX \DDX)^\AAA \to R \DDX \DDX$ strongly dinatural in $\DDX$.

We define this monad in the form of a Kleisli triple (see Manes~\cite{manes} under the name `clone forms', or Moggi~\cite[Definition 1.2]{DBLP:journals/iandc/Moggi91}), which consists of an assignment of objects~$\CR$, the unit of the monad $\eta_\AAA : \AAA \to \CR \AAA$, and lifting of morphisms $f : \AAA \to \CR \AAB$ to $f^* : \CR \AAA \to \CR \AAB$ subject to appropriate conditions. We prefer this format, because the usual `monoid' form of $\CR$ is quite cumbersome to work with in this case, as the argument $\AAA$ appears in a double-negative position.

First, we define the object assignment. For this, we note that for a morphism $f : \AAA \to \AAB$ in $\AA$, the collection of morphisms $\AA[f, R\DDX \DDX]$ indexed by~$\DDX$ is a strong dinatural transformation $\AA[\AAB, R{-}{=}] \to \AA[\AAA, R{-}{=}]$. Since strong dinatural transformations compose, this means that for all objects $\AAC$ in $\AA$, $\bdin[\AA[\AAC, R{-}{=}], \AA[\AAB, R{-}{=}]]$ is contravariantly functorial in $\AAB$ with the action on a morphism $f : \AAA \to \AAB$ given as follows:
  \begin{equation*}
   \theta \mapsto \AA[f, R{-}{=}] \cdot \theta : \bdin[\AA[\AAC, R{-}{=}], \AA[\AAB, R{-}{=}]] \to \bdin[\AA[\AAC, R{-}{=}], \AA[\AAA, R{-}{=}]]
  \end{equation*}
We use this functor in the definition of the object assignment  $\CR$:
\begin{definition}\label{d:cra}
Let $\AA$, $\DD$, and $R$ be as above. For an object $\AAA$ in $\AA$, we call an object $\CR \AAA$ in $\AA$ an \emph{$\AAA$-continuation} if there exists the following isomorphism natural in $\AAC$:
\begin{equation}\label{eq:cra}
  \AA[\AAC,\CR \AAA] \cong \bdin[\AA[\AAA, R{-}{=}], \AA[\AAC, R{-}{=}]]
\end{equation}
\end{definition}

Of course, such an object may or may not exist for a given $\AAA$, and if it does, it is defined up to isomorphism. In the remainder, we assume a particular choice of an $\AAA$-continuation for an object $\AAA$ (provided at least one exists), denoted~$\CR \AAA$.

We now spell out some details of Definition~\ref{d:cra}. We write $\tobdn{\spholder}$ for the isomorphism from left to right, that is, for an $\AA$-morphism $f : \AAC \to \CR \AAA$, we obtain a strong dinatural transformation $\tobdn{f} : \AA[\AAA, R{-}{=}] \to \AA[\AAC, R{-}{=}]$. In the other direction, for a strong dinatural transformation $\theta : \AA[\AAA, R{-}{=}] \to \AA[\AAC, R{-}{=}]$, we get $\tocay{\theta} : \AAC \to \CR \AAA$. Using this notation, the naturality conditions can be stated as follows:
\begin{itemize}
\item For morphisms $g : \AAB \to \CR A$ and $f : \AAC \to \AAB$ in $\AAA$, it is the case that $\AA[f, R{-}{=}] \cdot \tobdn{g} = \tobdn{g \cdot f}$,
\item For a strong dinatural transformation $\theta : \AA[\AAA, R{-}{=}] \to \AA[\AAC, R{-}{=}]$ and an $\AA$-morphism $f : \AAB \to \AAC$, it is the case that $\tocay{\theta} \cdot  f = \tocay{\AA[f, R{-}{=}] \cdot \theta}$.
\end{itemize}

\begin{remark}
  \label{rem:dinatHom}
Given a strong dinatural transformation $\theta : \AA[\AAA, R{-}{=}] \to \AA[\AAB, R{-}{=}]$, if we unfold the definition of the hom-functor, the condition for strong dinaturality of $\theta$ can be stated as follows: For all $f : \DDX \to \DDY$ in $\DD$, $g_1 : \AAA \to R \DDX \DDX$ and $g_2 : \AAA \to R \DDY \DDY$ in $\AA$, if $R \DDX f \cdot g_1 = Rf \DDY \cdot g_2$, then $R \DDX f \cdot \theta_{\DDX}(g_1) = Rf \DDY \cdot \theta_{\DDY}(g_2)$.
\end{remark}

Using the Yoneda lemma and the fact that currying and uncurrying with a constant preserves strong dinaturality, it is not difficult to show that if $\AA = \Set$ and $\CR \AAA$ exists for a set $\AAA$, then $\CR \AAA$ is (up to isomorphism) the set of transformations of the type $(R \DDX \DDX)^{\AAA} \to R \DDX \DDX$ strongly dinatural in $\DDX$.

We can now define the rest of the monad structure on $\CR$:

\begin{theorem}\label{thm:craMonad}
If $\CR \AAA$ exists for all objects $\AAA$ in $\AA$, the assignment $\CR$ can be given a monad structure defined by the unit $\eta$ and the Kleisli extension $\pholder^*$ as follows:
\begin{itemize}
\item $\eta_{\AAA} = \tocay{\id_{\AA[\AAA,R{-}{=}]}} : \AAA \to \CR \AAA$,
\item For $f : \AAA \to \CR \AAB$, we define $f^* = \tocay{\Phi^\AAA \cdot \tobdn{f}} : \CR \AAA \to \CR \AAB$, where $\Phi^\AAA = \tobdn{\id_{\CR \AAA}} : \AA[\AAA,R{-}{=}] \to \AA[\CR \AAA,R{-}{=}]$.
\end{itemize}
\end{theorem}

As the first example to illustrate this theorem, we consider the simplest setting: when $R$ is a constant $\Set$-bifunctor. In such a case, our construction instantiates to the usual continuation monad:

\begin{example}
We consider $\AA = \Set$ and write $\AAA \Rightarrow \AAB$ for the exponential object. Assume $RXY = O$ for a chosen set $O$. Since $R$ is constant, $\bdin[\Set[\AAA,R{-}{=}],\Set[\AAB,R{-}{=}]]$ is equal to $\Set[\Set[\AAA,O],\Set[\AAB,O]] = (\AAA \Rightarrow O) \Rightarrow (\AAB \Rightarrow O)$. Thus, $\CR \AAA$ is defined as $\CR \AAA = (\AAA \Rightarrow O) \Rightarrow O$ together with the isomorphism $\tobdn\spholder : \AAB \Rightarrow ((\AAA \Rightarrow O) \Rightarrow O) \cong (\AAA \Rightarrow O) \Rightarrow (\AAB \Rightarrow O) : \tocay{\spholder}$ given by $\tobdn{f}(k)(x) = f(x)(k)$ and $\tocay{g}(x)(k) = g (k) (x)$. We instantiate the construction from Theorem~\ref{thm:craMonad} with this data and obtain:
\begin{itemize}
\item $\eta(a) = k \mapsto k(a)$,
\item For $f : \AAA \to (\AAB \Rightarrow O) \Rightarrow O$, the extension $f^*(x) = k \mapsto x(y \mapsto f(y)(k))$.
\end{itemize}
One can compare this with the usual definition of the continuation monad via a Kleisli triple, e.g., in Moggi~\cite[Example 1.4]{DBLP:journals/iandc/Moggi91}.
\end{example}

\section{Codensity Monads}\label{s:cod}

Assume the categories $\AA$, $\DD$, and the bifunctor $R : \DD^\op \times \DD \to \AA$ as in the previous section. We say that $R$ is \emph{dummy} in its contravariant argument if it is of the shape $R \DDX \DDY = G \DDY$ for some $G : \DD \to \AA$.
In this section, we show that if $R$ is such a functor, the monad $\CR \AAA$ instantiates to the pointwise codensity monad of $G$.

In general, a \emph{codensity monad} of a functor $G$ (see, e.g., Mac Lane~\cite{catwork}) is the right Kan extension of $G$ along itself. Here, we are interested in \emph{pointwise} codensity monads, which can given by the following end formula:
\begin{equation}\label{eq:codcod}
  \mathfrak{T}^GA = \int_{X \in \DD} \AA[A, GX] \power GX
\end{equation}
The notation $\int_XFXX$ denotes the \emph{end} of a mixed-variant functor $F$. The symbol $\AA[\AAA, RX] \power RX$ denotes the $\AA[\AAA, RX]$-\emph{power} of $RX$, which can be thought of as an $\AA[\AAA, RX]$-ary product of $RX$ with itself, and can be concisely defined as an object $P$ that satisfies the following isomorphism natural in~$\AAC$:
\begin{equation}\label{eq:power}
 \AA[\AAC, P] \cong \Set[\AA[\AAA, RX], \AA[\AAC, RX]]
\end{equation}
(The similarity of~\eqref{eq:power} and~\eqref{eq:cra} is of course not coincidental.) We use the simple fact that strong dinaturality collapses to the usual naturality if the involved bifunctors are simply functors:

\begin{lem}\label{lem:dinatToNat}
Given two functors $G, H : \DD \to \AA$, we can treat them as
bifunctors $\DD^\op \times \DD \to \AA$ dummy in their first
argument. Then, strong dinatural transformations $G \to H$ are exactly
natural transformations $G \to H$.
\end{lem}

We use this lemma and some known facts about ends to show that the codensity monad of $G$ satisfies the isomorphism from Definition~\ref{d:cra}:
\begin{align*}
  &\ \bdin[\AA[\AAA, G{-}], \AA[\AAC, G{-}]] \\[1ex] 
  \cong &\ \mathbf{Nat}[\AA[\AAA, G{-}], \AA[\AAC, G{-}]] \tag{Lemma~\ref{lem:dinatToNat}} \\
  \cong &\ \int_\DDX \mathbf{Set}[\AA[\AAA, G\DDX], \AA[\AAC, G\DDX]] \tag{sets of nat. trans. as ends} \\
  \cong &\ \int_\DDX \AA[\AAC, \AA[\AAA, G\DDX] \power G\DDX] \tag{def. of power}\\
  \cong &\ \AA[\AAC, \int_\DDX \AA[\AAA, G\DDX] \power G\DDX] \tag{ends commute with hom-functors}\\
\end{align*}
Applying the Yoneda lemma and comparing the monad structures, we get:
\begin{theorem}
For a functor $R\DDX \DDY = G\DDY$ dummy in its first argument, assuming the mentioned ends exist, the monad $\CR$ exists and is isomorphic to the codensity monad of~$G$.
\end{theorem}

We also remark on other candidates for generalised codensity monads:

\begin{remark}\label{rem:differentMonads}
The proof of Theorem~\ref{thm:craMonad} relies on the naturality condition in the equation~\eqref{eq:cra}, but not explicitly on strong dinaturality. This suggests that we could define more monads by replacing $\bdin$ with different kinds of transformations. However, the other obvious choices do not lead to anything interesting. Transformations natural in both arguments yield the usual codensity monad of the bifuctor treated as a functor from the product category $\DD^\op \times \DD$, while dinatural transformations do not compose in general, so in such a case the construction in Definition~\ref{d:cra} does not yield a monad.
\end{remark}

\section{Dicodensity Monads from Limits}\label{sec:genCod}

Codensity monads can be alternatively constructed as a limit over a diagram indexed by a comma category. In this section, we show a similar result for our dicodensity monads, however, the diagram needs to be slightly more involved.

The codensity monad of a functor $G$ on an object $A$, that is, $\mathfrak{T}^G A$, can be given as the limit of the diagram $(A \downarrow G) \xrightarrow{\pi} \DD \xrightarrow{G} \AA$, where $A \downarrow G$ is the comma category and $\pi$ is the usual projection functor. To generalise this definition to mixed-variant bifunctors, we first define the following generalisation of the comma category:

\begin{definition}
  Given an object $A$ in a category $\AA$ and a bifunctor $R : \DD^\op \times \DD \to \AA$ for a category $\DD$, we define the category $A \Downarrow R$ as follows:
  \begin{itemize}
    \item Objects are pairs $(\DDX, f : A \to R \DDX \DDX)$, where $\DDX$ is an object in $\DD$,
    \item Arrows between objects $(\DDX, f)$ and $(\DDY, g)$ are morphisms $d : \DDX \to \DDY$ in $\DD$ such that the following diagram commutes in $\AA$:
          \begin{equation}
            \label{eq:dicodenseArr}
            \begin{tikzpicture}[-latex, scale=1, baseline={(current bounding box.center)}]
        \node (a) at (0,0) {$A$};
        \node (b0) at (-1,-1) {$R \DDX \DDX$};
        \node (b1) at (1,-1) {$R \DDY \DDY$};
        \node (c) at (0,-2) {$R\DDX\DDY$};
      \path[arrows={-latex}, font=\scriptsize]
      (a) edge node [left,xshift=-0.1em,yshift=0.3em] {$f$} (b0)
      (a) edge node [right,xshift=0.1em,yshift=0.3em] {$g$} (b1)
      (b0) edge node [left,xshift=-0.1em,yshift=-0.2em] {$\urep{\DDX}{d}$} (c)
      (b1) edge node [right,xshift=0.1em,yshift=-0.2em] {$\urep{d}{\DDY}$} (c)
    ;
  \end{tikzpicture}
          \end{equation}
  \end{itemize}
\end{definition}

Now, we do not take the limit of the projection composed with $R$ (in particular, they don't compose), because we need to include all the objects of the shape $R\DDX\DDY$ together with the morphisms $Rd\DDY$ and $R\DDX d$. In other words, we want to take the limit over all the bottom halves of the diagrams~\eqref{eq:dicodenseArr} combined. However, since they are not in the data of $A \Downarrow R$, we need to add one more layer to the construction of the diagram:

\newcommand{\bunting}{\triangledown}

\begin{definition}\label{d:buntinag}
Given $R$ and $A$ as above, we define the \emph{bunting category} of $A \Downarrow R$, denoted $(A \Downarrow R)^{\bunting}$, that consists of the following data:
\begin{itemize}
\item For each object $(\DDX,f)$ in $A \Downarrow R$, there is an object $(\DDX,f,R\DDX\DDX)$,
  \item For each arrow $d : (\DDX,f) \to (\DDY, g)$ in $A \Downarrow R$, there is an object $(d, R\DDX \DDY)$,
\item For each arrow $d$ as above, there are two morphisms $R \DDX d : (\DDX, f, R\DDX\DDX) \to (d, R\DDX \DDY)$ and $Rd\DDY : (\DDY, g, R\DDY\DDY) \to (d, R\DDX\DDY)$.
\end{itemize}
\end{definition}

There exists an obvious projection $\pi^\bunting : (A \Downarrow R)^{\bunting} \rightarrow \AA$ that selects the last components of objects, domains, and codomains.

\begin{theorem}
If the limit of $\pi^\bunting$ exists for all $A$, it is isomorphic to $\CR A$
\end{theorem}

This theorem follows easily from the observation that in the light of Remark~\ref{rem:dinatHom} strong dinatural transformations $\AA[A, R{-}{=}] \to \AA[\AAB, R{-}{=}]$ are in a natural 1-1 correspondence with cones over $\pi^\bunting$, while the morphisms $\AAB \to \CR A$ from~\eqref{eq:cra} are the mediators of the limit.

It follows that if this limit exists for all $A$, it is isomorphic to $A$ if and only if $\CR$ is identity -- a counterpart of the known result that codensity monad $\mathfrak{T}^G$ is identity if and only if $G$ is codense (that is, the limit of $(A \downarrow G) \xrightarrow{\pi} \DD \xrightarrow{G} \AA$ is isomorphic to $A$).

\section{Monads from Representations}\label{s:cayley}

In this section, given a bifunctor $R : \DD^\op \times \DD \to \AA$, we introduce a set of sufficient conditions for the monad $\CR$ on $\AA$ to be isomorphic to the monad $UF$ for an adjunction $F \dashv U$ with $F : \AA \to \CC$ for a category $\CC$. Broadly speaking, the conditions require that there exists a functor $\Ubar : \CC \to \DD$, and that for all objects $X$ in $\DD$, $RXX$ can be lifted to an object $\mathbb{R}X$ in~$\CC$, such that for all~$M$ in~$\CC$, the object~$\mathbb{R}(\Ubar M)$ is a form of a representation of~$M$.

Both the definition of the representation and the proof of the isomorphism theorem generalise our previous results~\cite{DBLP:conf/fossacs/PirogPS19}, where $\AA = \DD = \Set$: in that case, we simply take $\Ubar = U$. In the general case, the types of the involved functors are summarised in the following diagram:

\begin{equation}
  \label{eq:invovedCats}
\begin{tikzpicture}[-latex, scale=1.5]
    \node (a0) at (0,0) {$\DD^\op \times \DD$};
    \node (a1) at (2,0) {$\DD$};
    \node (b1) at (2,-1) {$\CC$};
    \node (b0) at (0,-1) {$\AA$};
    \node at (1,-1) {$\bot$};
    \path[arrows={-latex}, font=\scriptsize]
    (a0) edge node [left] {$R$} (b0)
    (b0) edge [bend left] node [auto] {$F$} (b1)
    (b1) edge [bend left] node [auto] {$U$} (b0)
    (b1) edge node [right] {$\Ubar$} (a1)
    ;
\end{tikzpicture}
\end{equation}

We first introduce some notation: given a morphism $f : A \to UX$, we write $f' : FA \to X$ for the contraposition of $f$ via the adjunction (the unique homomorphism induced by the freeness of $FA$), and use $\widehat f$ for $U f' : UFA \to UX$.

\begin{definition}\label{d:tight}
Let $F : \AA \to \CC$ be a left adjoint to
$U : \CC \to \AA$. A \emph{representation} consists of the
following components and conditions:
\begin{enumerate}
\item A bifunctor
  $\urepname : \catvar{D}^\op \times \catvar{D} \to \AA$,
\item For each object $X$ in $\catvar{D}$, an object $\repxx$ in $\CC$ such
  that:\begin{enumerate}\item $U\repxx = \urepxx$, \item the assignment $\ufreem{\spholder} : \AA[A,R{-}{=}] \to \AA[UFA,R{-}{=}]$ is strongly dinatural,
    \end{enumerate}
  \item A functor  $\Ubar : \CC \to \catvar{D}$ and for each object $M$ in $\CC$,   a morphism
    $\sigma_M : M \to \rep{(\Ubar M)}{(\Ubar M)}$ in~$\CC$ such that:\begin{enumerate}\item
    $U\sigma_M : UM \to \urep{(\Ubar M)}{(\Ubar M)}$ is strongly dinatural in $M$, \item $\sigma$ commutes with strong dinatural transformations, that is, for $\theta : \AA[A,R{-}{=}] \to \AA[B,R{-}{=}]$ and $k : A \to RXX$, the morphism $\theta_{\Ubar \mathbb R X} (U \sigma_{\mathbb R X} \cdot k)$ factors as:
    \begin{equation}\label{eq:factorisation}
      B \xrightarrow{\theta_X(k)} U \mathbb R X \xrightarrow{U\sigma_{\mathbb R X}} U \mathbb R (\Ubar \mathbb R X)
    \end{equation}
  \end{enumerate}
\item A strong dinatural transformation
    $\rho_M :  R (\Ubar M)(\Ubar M) \to UM$ such that
    $\rho_M \cdot U\sigma_M = \id$.
\end{enumerate}
\end{definition}

This definition is quite involved, so it might be useful to take a closer look at some of its components:

\begin{remark}\label{rem:reps}
  Unfolding the definition of hom-functor, the condition of $\ufreem\spholder$ being strongly dinatural can be stated more explicitly as follows.
  For an object $A$ in~$\AA$, objects $X$, $Y$ in~$\catvar{D}$, morphisms
  $f_1 : A \to \urepxx$, $f_2 : A \to \urepyy$ in $\AA$, and a morphism $g : X \to Y$ in~$\catvar{D}$, it is
  the case that
    \begin{equation*}
      \text{if}
      \quad
      \begin{tikzpicture}[-latex, scale=1, baseline={(current bounding box.center)}]
        \node (a) at (0,0) {$A$};
        \node (b0) at (1,1) {$\urepxx$};
        \node (b1) at (1,-1) {$\urepyy$};
        \node (c) at (2,0) {$\urep{X}{Y}$};
      \path[arrows={-latex}, font=\scriptsize]
      (a) edge node [left,yshift=0.3em] {$f_1$} (b0)
      (a) edge node [left,yshift=-0.3em] {$f_2$} (b1)
      (b0) edge node [right,yshift=0.3em] {$\urep{X}{g}$} (c)
      (b1) edge node [right,yshift=-0.3em] {$\urep{g}{Y}$} (c)
    ;
  \end{tikzpicture}
  \quad
  \text{commutes, then}
  \quad
        \begin{tikzpicture}[-latex, scale=1, baseline={(current bounding box.center)}]
        \node (a) at (0,0) {$UFA$};
        \node (b0) at (1,1) {$\urepxx$};
        \node (b1) at (1,-1) {$\urepyy$};
        \node (c) at (2,0) {$\urep{X}{Y}$};
      \path[arrows={-latex}, font=\scriptsize]
      (a) edge node [left,yshift=0.3em] {$\ufreem{f_1}$} (b0)
      (a) edge node [left,yshift=-0.3em] {$\ufreem{f_2}$} (b1)
      (b0) edge node [right,yshift=0.3em] {$\urep{X}{g}$} (c)
      (b1) edge node [right,yshift=-0.3em] {$\urep{g}{Y}$} (c)
    ;
  \end{tikzpicture}
  \quad
      \text{commutes.}
    \end{equation*}
  \end{remark}
  \begin{remark}
    \label{rem:repsb}
A sufficient condition for $\sigma$ to commute with strong dinatural transformations is the following: For each object $X$ in $\catvar{D}$, a set of indices $I_X$ and a
    family of morphisms $\appm_{X,i} : \Ubar\rep{X}{} \to X$, where
    $i \in I_X$, such that $\urep{(\Ubar\rep{X}{})}{\appm_{X}}$ is a
    jointly monic family, and the following diagram commutes for all $X$ and $i \in I_X$:
\begin{equation*}
\begin{tikzpicture}[-latex, scale=1.5]
    \node (a0) at (0.7,0) {$U\rep{X}{}$};
    \node (a1) at (3,0) {$\urep{(\Ubar\rep{X}{})}{(\Ubar\rep{X}{})}$};
    \node (b1) at (3,-1) {$\urep{(\Ubar\rep{X}{})}{X}$};
    \node (b0) at (0.7,-1) {$\urepxx$};
    \path[arrows={-latex}, font=\scriptsize]
    (a0) edge node [auto] {$U\sigma_\repxx$} (a1)
    (a1) edge node [auto] {$\urep{(\Ubar\rep{X}{})}{\appm_{X,i}}$} (b1)
    (b0) edge node [auto] {$\urep{\appm_{X,i}}{X}$} (b1)
    (a0) edge node [auto] {$\id$} (b0)
    ;
\end{tikzpicture}
\end{equation*}
\end{remark}

\begin{theorem}
  \label{thm:cayleyMonad}
Given a representation as in Definition~\ref{d:tight}, there is an isomorphism of monads $UF \cong \CR$.
\end{theorem}

In detail, this means that $UFA$ satisfies the isomorphism~\eqref{eq:cra} for all $A$ and that the two monadic structures coincide. In one direction, the isomorphism can be stated as follows: given $f : \AAC \to UFA$, we define $\tobdn{f}_X(k : A \to RXX) = \AAC \xrightarrow{f} UFA \xrightarrow{\ufreem{k}} RXX$. In the other direction, a strong dinatural transformation
$\theta : \AA[A, R{-}{=}] \to \AA[\AAC, R{-}{=}]$
corresponds to the morphism
$\tocay{\theta} = \AAC \xrightarrow {\theta_{\Ubar FA}(U\sigma_{FA} \cdot \eta_A)} \urep{(\Ubar FA)}{(\Ubar FA)} \xrightarrow{\rho_{FA}} UFA$.

We discuss two examples. The first one instantiates Theorem~\ref{thm:cayleyMonad} to the usual Cayley representation of monoids and lists, while the second one shows that the theorem can be used to obtain the known result that for an adjunction $F \dashv U$, the codensity monad $\mathfrak{T}^U$ is isomorphic to the monad $UF$.

\begin{example}\label{ex:listOnSet}
Consider the adjunction $F \dashv U$ with $F : \Set \to \catname{Mon}$ being the usual free monoid functor. We use Theorem~\ref{thm:cayleyMonad} to show that $\mathfrak{C}^{\Set[{-}, {=}]}$ is isomorphic to the monad~$UF$ (the list monad). We show that $F \dashv U$ satisfies Definiton~\ref{d:tight}:
  \begin{enumerate}
  \item The categories are $\AA = \DD = \Set$, $\CC = \catname{Mon}$, and the bifunctor is the hom-functor $\Set[{-}, {=}]$, that is, $RXY = Y^X$,
  \item $\mathbb R X = (X^X,\ \circ, \ \id)$. The fact that it is a monoid is known from the Cayley theorem for monoids,
    \begin{enumerate}
    \item Trivially, $U \mathbb R X = X^X = RXX$,
    \item To show that $\widehat{-}$ is strongly dinatural, we use Remark~\ref{rem:reps}: given a set $A$ and some elements $a,b, \ldots, c \in A$, we need to see
  the following:
  \begin{equation*}g \circ f_1(a) \circ f_1(b) \circ \cdots \circ f_1(c) = f_2(a)
  \circ f_2(b) \circ \cdots \circ f_2(c) \circ g
\end{equation*}
Fortunately, the
  assumption, which in this case becomes
  $g\circ f_1(a) = f_2(a) \circ g$ for all $a \in A$, allows us to
  ``commute''~$g$ from one side of the chain of function compositions
  to the other.
    \end{enumerate}
  \item We define $\Ubar = U$ and $\sigma_{(M,\cdot,\varepsilon)}(a) = b \mapsto a \cdot b$. The fact that $\sigma$ is a homomorphism is known from the Cayley theorem,
    \begin{enumerate}
    \item To show that $\sigma$ is strongly dinatural, let $M$ and $N$ be two monoids, and let $f : M \to N$ be a homomorphism such that $f(m) = n$ for some $m \in M$ and $n \in N$. We calculate:
      \begin{align*}
             RMf(\sigma_M(m))
           =\ &   RMf(b \mapsto m \cdot b) \tag{def. of $\sigma$}\\
         =\ &     b \mapsto f(m \cdot b) \tag{def. of $R$.}\\
        =\ &      b \mapsto f(m) \cdot f(b) \tag{$f$ is a homomorphism}\\
       =\ &       b\mapsto n \cdot f(b) \tag{assumption above}\\
     =\ &         RfN(b \mapsto n \cdot b) \tag{def. of $R$.}\\
    =\ &          RfN(\sigma_N(n)) \tag{def. of $\sigma$}
      \end{align*}
    \item To show that $\sigma$ commutes with strong dinaturals, we use Remark~\ref{rem:repsb}. We define $I_X = X$ and $\mathsf{run}_{X,i}(f) = f(i)$. We instantiate the diagram with $RXY = X \Rightarrow Y$, and show that it commutes for all $i \in X$, point-wise for all arguments if applied to $f : X \to X$ and then the result to $f' : X \to X$:
      \begin{align*}
        & (  ((X \Rightarrow X) \Rightarrow \mathsf{run}_{X,i}) \circ U\sigma_{(X \Rightarrow X, \circ, \mathsf{id})}  )(f)(f') \\
        =\ & (((X \Rightarrow X) \Rightarrow \mathsf{run}_{X,i})(U\sigma_{(X \Rightarrow X, \circ, \mathsf{id})}(f)))(f') \tag{composition} \\
          =\ & (((X \Rightarrow X) \Rightarrow \mathsf{run}_{X,i})(g \mapsto f \circ g))(f') \tag{def. of $\sigma$} \\
        =\ & (\mathsf{run}_{X,i} \circ (g \mapsto f \circ g))(f') \tag{hom-functor} \\
        =\ & \mathsf{run}_{X,i} (f \circ f') \tag{composition} \\
        =\ & f (f'(i)) \tag{def. of $\mathsf{run}$} \\   
        =\ & f (\mathsf{run}_{X,i} (f')) \tag{def. of $\mathsf{run}$} \\
        =\ & (f \circ \mathsf{run}_{X,i}) (f') \tag{composition} \\
        =\ & (\mathsf{run}_{X,i} \Rightarrow X)(f) (f') \tag{hom-functor} \\
      \end{align*}
  
    \end{enumerate}
  \item We define $\rho_{(M,\cdot,\varepsilon)}(g) = g(\varepsilon)$. To show that it is strongly dinatural, assume two monoids $M$ and $N$, together with $g_1 \in M^M$, $g_2 \in N^N$, and a homomorphism $f : M \to N$ such that $RfM(g_1) = RNf(g_2)$. We calculate:
      \begin{align*}
        f(\rho_M(g_1)) =\ & f(g_1(\varepsilon_M)) \tag{def. of $\rho$} \\
        =\ &  RfM(g_1)(\varepsilon_M) \tag{def. of $R$} \\
        =\ & RNf(g_2)(\varepsilon_M) \tag{assumption above}\\
        =\ & g_2(f(\varepsilon_M)) \tag{def. of $R$} \\
           =\ &    g_2(\varepsilon_N) \tag{$f$ is a homomorphism}
      \end{align*}    
\end{enumerate}
\end{example}

The case for the usual Cayley representation of monoids and lists can also be trivially obtained by instantiating Theorem~\ref{thm:commutativeCCC} given in Section~\ref{s:homo} with $M = \mathsf{\Id}$ (the identity monad).
The next example shows that Theorem~\ref{thm:cayleyMonad} is general enough to cover also the well-known case of the codensity monad of a right adjoint:

\begin{example}
Let $U : \CC \to \AA$ be a functor with a left adjoint $F$. Let $\DD = \CC$ and $RXY = UY$. We use Theorem~\ref{thm:cayleyMonad} to show that $\mathfrak{C}^R$ and $UF$ are isomorphic as monads. We take $\Ubar = \mathsf{Id}$, and $\mathbb{R}X = X$. The condition (2a) is trivial. To show~(2b), we use Remark~\ref{rem:reps}: We assume that
\begin{tikzpicture}[-latex, scale=1, baseline={(a.base)}]
        \node (a) at (-0.5,0) {$A$};
        \node (b0) at (1,0.45) {$UX$};
        \node (b1) at (1,-0.45) {$UY$};
      \path[arrows={-latex}, font=\scriptsize]
      (a) edge node [above,yshift=0.0em] {$f_1$} (b0)
      (a) edge node [below,yshift=-0.0em] {$f_2$} (b1)
      (b0) edge node [right,yshift=0.0em] {$Ug$} (b1) 
    ;
  \end{tikzpicture} commutes.
  Its contraposition via the adjunction is
  \begin{tikzpicture}[-latex, scale=1, baseline={(a.base)}]
        \node (a) at (-0.5,0) {$FA$};
        \node (b0) at (1,0.45) {$X$};
        \node (b1) at (1,-0.45) {$Y$};
      \path[arrows={-latex}, font=\scriptsize]
      (a) edge node [above,yshift=0.0em] {$f_1'$} (b0)
      (a) edge node [below,yshift=-0.0em] {$f_2'$} (b1)
      (b0) edge node [right,yshift=0.0em] {$g$} (b1)
    ;
  \end{tikzpicture}%
 , and its $U$-image is the desired diagram~\begin{tikzpicture}[-latex, scale=1, baseline={(a.base)}]
        \node (a) at (-0.5,0) {$UFA$};
        \node (b0) at (1,0.45) {$UX$};
        \node (b1) at (1,-0.45) {$UY$};
      \path[arrows={-latex}, font=\scriptsize]
      (a) edge node [above,xshift=-0.99em] {$\ufreem{f_1} = Uf_1'$} (b0)
      (a) edge node [below,xshift=-0.99em] {$\ufreem{f_2} = Uf_2'$} (b1)
      (b0) edge node [right,yshift=0.0em] {$Ug$} (b1)
    ;
  \end{tikzpicture}.
  For (3), we note that $\mathbb{R}(\Ubar M) = M$. We define $\sigma = \id$, which makes (3a) and (3b) trivial. For (4), we note that $R(\Ubar M)(\Ubar M) = UM$, and define $\rho = \id$, which is trivially a retraction of $U\sigma$.
\end{example}

\section{Monads Induced by Objects of Homomorphisms}\label{s:homo}

\newcommand\sten{{\otimes}}
\newcommand\sfun{{\Rrightarrow}}
\newcommand\hereApp{\mathsf{happ}}
\newcommand\icomp{\mathsf{hcomp}}
\newcommand\iid{\mathsf{hid}}

While the notion of representation introduced in Definition~\ref{d:tight} is powerful enough to characterise a class of dicodensity monads, it is rather difficult to use due to the number and complexity of technical conditions. In this section, we explore a particular subclass of representations, which generalises Example~\ref{ex:listOnSet} from functions on sets to internalised morphisms between algebras of a commutative monad.
 In such a setting, we can substantially simplify the required conditions. First, consider two representative examples:

\begin{example}\label{ex:global}
  In programming, the list monad can be used to model nondeterministic computations, which produce a number of possible outcomes in a specific order. The list monad can be extended to provide `global error': if an error occurs in any branch, the entire computation fails. First, we consider the $A \mapsto A + 1$ monad, known in programming as the \emph{maybe} monad, which arises from the adjunction between $\Set$ and $\Set_\bullet$  (the category of pointed sets and point-preserving maps). The `global error' monad, $\flift \MM A = \mathcal{L}A + 1$, where $\mathcal L$ is the list monad, is induced by the following  distributive law of the list monad over maybe:
  \begin{equation*}
  \begin{cases}
    [\inl\,x_1, \ldots, \inl\,x_n] = \inl\,[x_1, \ldots, x_n] & \text{if all elements of the list are of the form $\inl\,x$} \\
     [\ldots, \inr\,*, \ldots] = \inr\,* & \text{if at least one element of the list is of the form $\inr\,*$}
\end{cases}
\end{equation*}
This monad is isomorphic to $\CR$ via Theorem~\ref{thm:cayleyMonad} with $\AA =\Set$, $\CC = \flift \MM\Alg$ (the category of algebras for~$\flift \MM$), $\DD = \Set_\bullet$, and $\Ubar(A,a) = (A, a(\inr\,*))$. We define $R(X,x)(Y,y) = \{ f : X\to Y \ |\ f(x) = y \}$ (that is, $R$ is the hom-functor of~$\Set_\bullet$), and $\mathbb R (X,x) = (R(X,x)(X,x), a)$, where:
\begin{equation*}
\begin{cases}
    a([\inl\,f_1, \ldots, \inl\,f_n]) = f_1 \circ \cdots \circ f_n & \text{if all elements of the list are of the form $\inl\,f$} \\
     a([\ldots, \inr\,*, \ldots]) = \mathsf{const}(x) & \text{if at least one element of the list is of the form $\inr\,*$}
\end{cases}
\end{equation*}
\end{example}

\begin{example}\label{ex:semi}
  Let $\AA = \Set$ and $\CC = \catname{ISRing}$ (the category of idempotent semirings). The monad that arises from free idempotent semirings is the composition $\mathcal{F}\mathcal{L}$, where $\mathcal F$ is the finite powerset monad, and $\mathcal L$ is the list monad. The monad structure comes from the distributive law between monads $[X_1,\ldots,X_n] \mapsto \{[x_1,\ldots,x_n] \ |\ x_i \in X_i \}$. We can construct the representation of idempotent semirings using Theorem~\ref{thm:cayleyMonad} assuming $\DD$ to be $\catname{ICMon}$ (the category of idempotent commutative monoids, that is, pointed semilattices), $\Ubar : \catname{ISRing} \to \catname{ICMon}$ to be the functor that forgets the multiplicative structure, $R$ to be the set of homomorphisms between idempotent commutative monoids, that is, the hom-functor $RXY = \catname{ICMon}[X,Y] : \catname{ICMon}^\op \times \catname{ICMon} \to \Set$, and the idempotent semiring $\mathbb R X$ given by the composition of homomorphisms (as in the usual Cayley representation) for the multiplicative part, and the additive structure defined pointwise.
\end{example}

While these examples are covered by Theorem~\ref{thm:cayleyMonad}, it requires manually checking a number of complex conditions. Instead, we consider the case when the category $\DD$ is the category of $\MM$-algebras for a commutative monad $\MM$, and $RXY$ represents $\MM$-homomorphisms between $X$ and $Y$. This allows us to simplify the structure required for the theorem to hold.

\subsection{Objects of homomorphisms}

First, we recall the definition of objects of homomorphisms. Let $(\AA, \otimes, I)$ be a monoidal category, and $\MM$ be a bistrong monad on $\AA$ with strengths $\tau_{A,B} : \MMA \otimes B \to \MM(A \otimes B)$ and $\tau'_{A,B} : A \otimes \MMB \to \MM(A \otimes B)$.
\begin{definition}(Kock~\cite{Kock_1971})
  \label{def:homobjs}
  A morphism $h : X \otimes A \to B$ in $\AA$ is \emph{$1$-linear}
  with respect to $\MM$-algebras $(A,a)$ and $(B,b)$ if the following
  diagram commutes:
\begin{equation*}
  \begin{tikzpicture}[scale=0.75, baseline={(current bounding box.center)}]
\node (a) at (0,0) {$X \sten \MMA$};
\node (b) at (4,0) {$\MM(X \sten A)$};
\node (c) at (8,0) {$\MMB$};
\node (aa) at (0,-2) {$X \sten A$};
\node (cc) at (8,-2) {$B$};
\path[arrows={-latex}, font=\scriptsize]
(a) edge node [auto] {$\tau'$} (b)
(b) edge node [auto] {$\MM h$} (c)
(c) edge node [auto] {$b$} (cc)
(a) edge node [auto] {$X \sten a$} (aa)
(aa) edge node [auto] {$h$} (cc)
;
  \end{tikzpicture}
\end{equation*}
\end{definition}We write $\mathsf{1Lin}[X \otimes (A,a),\ (B,b)]$ for the set of all morphisms $X\otimes A \to B$ that are $1$-linear with respect to $(A,a)$ and $(B,b)$. It is straightforward to turn $\mathsf{1Lin}$ into a functor $\AA^\op \times \MM\Alg^\op \times \MM\Alg \to \Set$, which is used in the following definition:
\begin{definition}(Kock~\cite{Kock71closedcategories})
An \emph{object of homomorphisms} between $\MM$-algebras $(A,a)$ and $(B,b)$ is an object $(A,a) \Rrightarrow (B,b)$ that satisfies the following isomorphism natural in $X$:
\begin{equation}\label{eq:homobj}
\mathsf{1Lin}[X \otimes (A,a),\ (B,b)] \cong \AA[X, (A,a) \Rrightarrow (B,b)]
\end{equation}
\end{definition}
The Yoneda lemma gives us that if an object of homomorphisms exists for given algebras, it is unique up to isomorphism. Moreover, if $\Rrightarrow$ exists for all algebras, it can be given an obvious structure of a bifunctor inherited from~$\mathsf{1Lin}$.

\begin{remark}
  \label{rem:equaliser}
  A sufficient condition for the object of homomorphisms to exist is that $\AA$ has equalisers and is right-closed, that is, there exists the right adjoint:
  \begin{equation}
    \label{eq:rclosed}
\AA[A \otimes B, C] \cong \AA[A, B \Rightarrow C]
  \end{equation}
Concretely, $(A,a) \Rrightarrow (B,b)$ is given as the following equaliser
\begin{equation}
  \label{eq:equaliser}
\begin{tikzpicture}[scale=0.9, baseline={(current bounding box.center)}]
\node (a) at (0,0) {$(A,b) \sfun (B,b)$};
\node (b) at (3,0) {$A {\Rightarrow} B$};
\node (c) at (6.1,0.34) {$\MMA {\Rightarrow} \MMB$};
\node (d) at (9.5,0) {$\MMA {\Rightarrow} B$};
\path[arrows={-latex}, font=\scriptsize]
(a) edge (b)
(b) edge[bend left=4.9] node [above] {$\varphi$} (c)
(c) edge[bend left=5] node [above] {$\MMA {\Rightarrow} b$} (d)
(b) edge[bend right=10] node [below] {$a {\Rightarrow} B$} (d)
;
\end{tikzpicture}
\end{equation}
where $\varphi$ is the transpose of
$(A {\Rightarrow} B) \otimes \MMA \xrightarrow{\tau'} \MM((A {\Rightarrow} B) \otimes A) \xrightarrow{\MM\mathsf{app}} \MMB$ via~\eqref{eq:rclosed}.
\end{remark}

We use~\eqref{eq:homobj} to define the internal identity and composition morphisms analogous to the usual ones in closed monoidal categories:
\begin{itemize}
\item $\hereApp : ((A,a)\Rrightarrow (B,b)) \otimes A \to B$ as the transpose of $(A,a) \Rrightarrow (B,b) \xrightarrow{\id} (A,a) \Rrightarrow (B,b)$,
\item $\iid : I \to (A,a) \Rrightarrow (A,a)$ as the transpose of $I \otimes A \xrightarrow{\cong} A$,
\item
  $\icomp : ((B,b) \Rrightarrow (C,c)) \otimes ((A,a) \Rrightarrow (B,b)) \to (A,a) \Rrightarrow (B,b)$ as the transpose of:
\begin{align*}& (((B,b) \Rrightarrow (C,c)) \otimes ((A,a) \Rrightarrow (B,b))) \otimes A \xrightarrow{\cong} ((B,b) \Rrightarrow (C,c)) \otimes (((A,a) \Rrightarrow (B,b)) \otimes A) 
\\ & \qquad \xrightarrow{\id \otimes \hereApp} ((B,b) \Rrightarrow (C,c)) \otimes B \xrightarrow{\hereApp} C
  \end{align*}
\end{itemize}

\subsection{Cayley monads}

Assume that $\AA$ is a monoidal category. By $\mathbf{Mon}_\AA$ we denote the category of monoids in $\AA$, while $U^{\mathbf{Mon}} : \mathbf{Mon}_\AA \to \AA$ denotes the forgetful functor. Let $\MM$ be a commutative monad on~$\AA$, which means that the following diagram commutes for all~$A$ and~$B$, where $\tau$ and $\tau'$ denote the right and left strengths respectively:
\begin{equation*}
  \begin{tikzpicture}[scale=0.75]
\node (a) at (0,0) {$\MMA \otimes \MMB$};
\node (c) at (5,0) {$\MM(A \otimes \MMB)$};
\node (d) at (10,0) {$\MM\MM(A \otimes B)$};
\node (aa) at (0,-2) {$\MM(\MMA \otimes B)$};
\node (cc) at (5,-2) {$\MM\MM(A \otimes B)$};
\node (dd) at (10,-2) {$\MM(A \otimes B)$};
\path[arrows={-latex}, font=\scriptsize]
(a) edge node [auto] {$\tau$} (c)
(c) edge node [auto] {$\MM\tau'$} (d)
(d) edge node [auto] {$\mu$} (dd)
(a) edge node [auto] {$\tau'$} (aa)
(aa) edge node [auto] {$\MM\tau$} (cc)
(cc) edge node [auto] {$\mu$} (dd)
;
  \end{tikzpicture}
\end{equation*}
If the forgetful functor $U^\Mon$ is monadic, there exists a canonical distributive law between monads $\lambda : U^\Mon F^\Mon M \to \MM U^\Mon F^\Mon$~\cite{article}, where $F^\Mon$ is the left adjoint to $U^\Mon$. We denote the monad resulting from $\lambda$ as $\flift \MM$. We show that, with some additional assumptions, $\flift \MM$ is isomorphic to $\CR$ for $R$ assigning to a pair of $\MM$-algebras the object of homomorphisms between them:

\begin{theorem}
  \label{thm:commutativeCCC}
Let $\AA$ be a biclosed cocomplete moinoidal category and $M$ a commutative monad on $\AA$ such that:
\begin{itemize}
  \item $\AA$ admits homomorphism objects for $\MM$,
      \item $\AA$ is well-pointed (that is, the monoidal unit $I$ is a separator, or, equivalently, $\AA[I,\spholder]$ is faithful),
\end{itemize}
Let $\flift \MM = \MM U^\Mon F^\Mon$ be the composition of $\MM$ with the list monad via the canonical distributive law. Then, $\flift \MM$ is isomorphic to $\CR$ for $R : \MM\Alg^\op \times \MM\Alg \to \AA$ given by:
\begin{equation*}
R(A,a)(B,b) = (A,a) \Rrightarrow (B,b)
\end{equation*}
\end{theorem}

With the assumptions of Theorem~\ref{thm:commutativeCCC}, the functor $U^\Mon$ is indeed monadic, and the algebras of the monad $\flift \MM$ can be given, following~\cite[Section~6.2]{DBLP:journals/corr/Pirog16}, as tuples $(A, a, m, u)$, where $(A,a)$ is an $\MM$-algebra, and $(A,m,u)$ is a monoid in $\AA$ such that the following coherence condition holds:
\begin{equation*} 
  \begin{tikzpicture}[scale=0.75,baseline={(current bounding box.center)}]
\node (a) at (0,0) {$\MMA \otimes \MMA$};
\node (d) at (16,0) {$A\otimes A$};
\node (aa) at (0,-2) {$\MM(A\otimes MA)$};
\node (bb) at (5,-2) {$\MM\MM(A\otimes A)$};
\node (cc) at (10,-2) {$\MM\MMA$};
\node (ccc) at (13.25,-2) {$\MMA$};
\node (dd) at (16,-2) {$A$};
\path[arrows={-latex}, font=\scriptsize]
(a) edge node [auto] {$a \times a$} (d)
(d) edge node [auto] {$m$} (dd)
(a) edge node [auto] {$\tau$} (aa)
(aa) edge node [auto] {$M\tau'$} (bb)
(bb) edge node [auto] {$\MM\MM a$} (cc)
(cc) edge node [auto] {$\mu$} (ccc)
(ccc) edge node [auto] {$a$} (dd)
;
  \end{tikzpicture}
\end{equation*}
By $\Ubar : \flift \MM\Alg \to \MM\Alg$ we denote the functor that forgets the monoid part.

Theorem~\ref{thm:commutativeCCC} can be proved using Theorem~\ref{thm:cayleyMonad}, in which case the diagram~\eqref{eq:invovedCats} of the involved categories and functors becomes as follows:
\begin{equation*}
\begin{tikzpicture}[-latex, scale=1.5]
    \node (a0) at (0,0) {$\MM\text{-}\mathrm{Alg}^\op \times \MM\text{-}\mathrm{Alg}$};
    \node (a1) at (2,0) {$\MM\text{-}\mathrm{Alg}$};
    \node (b1) at (2,-1) {$\flift \MM\text{-}\mathrm{Alg}$};
    \node (b0) at (0,-1) {$\AA$};
    \node at (0.88,-1) {$\bot$};
    \path[arrows={-latex}, font=\scriptsize]
    (a0) edge node [left] {$R$} (b0)
    (b0) edge [bend left] node [above] {$F$} (b1)
    (b1) edge [bend left] node [below] {$U$} (b0)
    (b1) edge node [right] {$\Ubar$} (a1)
    ;
\end{tikzpicture}
\end{equation*}
The rest of the representation is defined as follows:
\begin{itemize}
  \item $\mathbb R (A,a) = ((A,a) \Rrightarrow (A,a), s , \icomp, \iid)$, where $s : \MM((A,a) \Rrightarrow (A,a)) \to ((A,a) \Rrightarrow (A,a))$ is the transpose via~\eqref{eq:homobj} of
    \begin{equation}\label{eq:s}
\MM((A,a) \Rrightarrow (A,a)) \otimes A
\xrightarrow{\tau}
\MM(((A,a) \Rrightarrow (A,a)) \otimes A)
\xrightarrow{\MM\hereApp}
\MMA
\xrightarrow{a}
A
\end{equation}
\item $\sigma_{(A,a,m,u)} : A \to (A,a) \Rrightarrow (A,a)$ is the transpose via~\eqref{eq:homobj} of $m$,
\item $\rho_{(A,a,m,u)} = ((A,a)\Rrightarrow (A,a)) \xrightarrow{\cong}  ((A,a)\Rrightarrow (A,a)) \otimes I \xrightarrow{\id \otimes u} ((A,a) \Rrightarrow (A,a)) \otimes A \xrightarrow{\hereApp} A$
\end{itemize}

The intuition behind the assumption that $\AA$ is cocomplete and biclosed is that these are sufficient conditions for $U^\Mon F^\Mon A$ to be the initial $I + (A \oplus \spholder)$-algebra. This gives us a more concrete characterisation of the $\ufreem{\spholder}$ operator in terms of a fold of the structure of the initial algebra, which seems needed for the condition (ii.b) of Theorem~\ref{thm:cayleyMonad} to hold. In general, strong dinatural transformations often appear in the literature in connection with the inital-algebra semantics of data structures~\cite{neumann2023paranaturalcategorytheory,DBLP:conf/fics/Uustalu10}.

The assumption that $\AA$ is well-pointed is used to show the condition (iii.b) of Theorem~\ref{thm:cayleyMonad} via Remark~\ref{rem:repsb}. The family $\run_{(A,a)}$ is indexed by the generalised elements $p : I \to A$, and is defined as:
\begin{equation*}
  \run_{(A,a),p} = (A,a) \sfun (A,a) \xrightarrow{\cong} ((A,a) \sfun (A,a)) \sten I \xrightarrow{\id \sten p} ((A,a) \sfun (A,a)) \sten A \xrightarrow{\hereApp} A
\end{equation*}
Then, the fact that the family $\urep{(\Ubar\rep{(A,a)}{})}{\appm_{(A,a)}}$ is jointly monic follows from the well-pointedness of $\AA$.

\begin{example}
In Example~\ref{ex:semi}, the monad $\MM$ is $\mathcal{F}$ (the finite powerset monad), whose algebras are idempotent commutative monoids. This means that $\flift \MM \Alg$ is the category of algebras $(A, \vee, \bot, \cdot, \varepsilon)$ such that $(A, \vee, \bot)$ is an idempotent commutative monoid, while $(A, \cdot, \varepsilon)$ is a monoid. The two need to satisfy $x(y \vee z) = xy \vee xz$, $(x \vee y)z = xz \vee yz$, and $x\bot = \bot x = \bot$, which makes $\flift \MM \Alg$ the category of idempotent semirings.
The value $R(A,\vee^A,\bot^A)(B,\vee^B,\bot^B) = (A,a) \Rrightarrow (B,a)$ becomes $\{ f : A \to B \ |\ f\text{ is a homomorphism } (A,\vee^A,\bot^A) \to (B,\vee^B,\bot^B)\}$, while $\mathbb{R}(A, \vee, \bot) = ( \{ f : A \to A \ | \ f \text{ is a homomorphism } (A, \vee, \bot) \to (A, \vee, \bot) \}, \vee^s, \bot^s, \circ, \id)$, where $(f \vee^s g)(a) = f(a) \vee g(a)$ and $\bot^s = a \mapsto \bot$.
\end{example}

\section{Conclusion and Future Work}\label{sec:conc}

This paper continues the exploration of continuation monads and Cayley representations in semantics, stemming from the work of Hinze~\cite{DBLP:conf/mpc/Hinze12}, and building on the formal footing provided by Paré and Román~\cite{PARE199833}. In this line, we build on our prior work~\cite{DBLP:conf/fossacs/PirogPS19}, where we only considered polynomial Cayley representations on $\Set$ and introduce generalised codensity monads over arbitrary bifunctors. We also relate them to generalised notion of representation, in particular where the bifunctors in question represent sets (or objects) of $\MM$-algebra homomorphisms, which is a common source of concrete examples.

One direction of future work would be to use the machinery developed in this paper to extend the polynomial representations over one variable to the multi-variable case, and explore the effects that they denote. Another possible question would be whether codensity liftings~\cite{DBLP:conf/calco/KatsumataS15} or similar constructions can be naturally generalised to mixed-variant bifunctors and whether such constructions have practical use in programming or semantics.

\section*{Acknowledgements}

We would like to thank Tarmo Uustalu and Dylan McDermott for useful discussions on strong dinatural transformations, and the anonymous reviewers, who provided very insightful comments. We are especially grateful to the reviewer who suggested significant technical improvements to Sections~\ref{sec:genCod} and~\ref{s:homo}.

\bibliographystyle{./entics}
\bibliography{dicodensity}

\end{document}